\documentclass[twocolumn,amsmath,aps,superscriptaddress,showpacs,floatfix,a4paper]{revtex4}
\usepackage{amsmath} 
\usepackage{graphicx} 
\usepackage{epsf}
\usepackage{bm} 
\usepackage{latexsym} 
\RequirePackage{amsopn}
\RequirePackage{amssymb} 
\RequirePackage{amsfonts}
\RequirePackage{amsthm} 
\usepackage[]{graphicx} 
\usepackage[]{amsmath}

\newcommand{\lb }{{\langle}}
\newcommand{\rb}{{\rangle}}     

\begin{document} 

\title{The crossover from single file to Fickian diffusion }
\author{Jimaan San\'e}
\affiliation{Rudolf Peierls Centre for Theoretical Physics,
           1 Keble Road, Oxford OX1 3NP, United Kingdom}
\affiliation{Department of Chemistry, Cambridge University, 
           Lensfield Road, Cambridge CB2 1EW, United Kingdom}
\author{Johan T.\ Padding}
\affiliation{Computational Biophysics, University of Twente, 
           PO Box 217, 7500 AE, Enschede, The Netherlands}
\author{Ard A.\ Louis}
\affiliation{Rudolf Peierls Centre for Theoretical Physics,
           1 Keble Road, Oxford OX1 3NP, United Kingdom}
\date{\today}

\begin{abstract}

The crossover from single-file diffusion, where the mean-square displacement scales as $\lb x^2\rb \sim t^{\frac12}$, to normal Fickian diffusion, where $\lb x^2\rb \sim t$, is studied as a function of channel width for colloidal particles. 
By comparing  Brownian dynamics to a hybrid molecular dynamics and mesoscopic simulation technique, we can study the effect of  hydrodynamic interactions on the single file mobility and on the crossover to Fickian diffusion for wider channel widths.  For disc-like particles with a steep interparticle repulsion, the single file mobilities for different particle densities are well described by  the exactly solvable  hard-rod model. This holds both for simulations that include hydrodynamics, as well as for those that don't.  When the single file constraint is lifted,  then for particles of diameter $\sigma$ and  pipe of width $L$ such that $(L- 2 \sigma)/\sigma   = \delta_c \ll 1$ the particles can be described  as hopping past one-another in an average time $t_{hop}$.  For shorter times $ t \ll t_{hop}$  the particles still exhibit sub-diffusive behaviour, but at longer times $t \gg t_{hop}$,  normal Fickian diffusion sets in with an effective diffusion constant $D_{hop} \sim 1/\sqrt{t_{hop}}$.
For the Brownian particles, $t_{hop} \sim \delta_c^{-2}$ when $\delta \ll 1$, but when hydrodynamic interactions are included,  we find a stronger dependence than $\delta_c^{-2}$. We attribute this difference to short-range lubrication forces that make it more difficult for particles to hop past each other in very narrow channels.
\end{abstract}

\maketitle

\section{Introduction}

When particles are confined to channels so narrow that mutual passage is excluded, the
geometric constraints restrict the particles to a single file and a fixed spatial sequence.
For short times the mean-square displacement may still take the Fickian form $\lb x^2 \rb = 2 D_0 t$, with a self-diffusion coefficient $D_0$, but at longer times the motion is strongly suppressed by collisions with neighbouring particles,  leading to an asymptotic scaling of the form:
\begin{equation}\label{eq:sfdmsd} 
\lb x^{2}\rb  = 2Ft^{1/2} 
\end{equation}
first derived by Harris~\cite{Harr65} in a more mathematical context, and independently by Levitt~\cite{Levi73} for particles  diffusing in a narrow pore.  Here $F$ is the single file diffusion (SFD) mobility.

There has been an increasing interest in transport through highly confined pores~\cite{Bura09}, stimulated 
in part by biological realisations such as ion channels~\cite{Mack04}, and aquaporins~\cite{Agre04}. Water under extreme nanoscale confinement exhibits behaviour that differs markedly from the bulk~\cite{Alle02}.  
Single file flow of water is also important in artificial materials like carbon nanotubes~\cite{Humm01,Fang08}.
Transport of simple molecules through porous materials such as zeolites also show single file sub-diffusive behaviour~\cite{Karg94,Hahn96,Smit08}.   

Although the biological and synthetic nanoscale systems described above show signatures of SFD, their interpretation is complicated by numerous other factors such as the interaction of the particles with the walls of the confining pore. By contrast well defined model systems can be created with micron sized colloidal particles.  One of the major advantages is that the particles can be directly  imaged in real time with digital video microscopy. By using  lithography~\cite{Wei00,Lin02,Lin05} or optical tweezers~\cite{Lut04}  to create the one-dimensional confinement for colloidal particles,  unambiguous evidence of  asymptotic SFD $\lb x^2 \rb \sim t^{\frac12}$ scaling was observed.    Lin {\em et al.}~\cite{Lin05} measured the  SFD mobility $F$ for different one-dimensional packing fractions $\eta = \rho \sigma$, where the density $\rho = N/L_p$,  the number of particles is $N$, the length of the pipe is $L_p$, and
 $\sigma$ is the colloidal hard-sphere radius.  They found good agreement between their measured $F$ and the SFD mobility for a hard-rod fluid (also known as a Tonks gas)~\cite{Levi73}:
\begin{equation}\label{eq:mobhr} 
F^{HR} = l_c\sqrt{\frac{D_{o}}{\pi}} =
\frac{\sigma(1-\eta)}{\eta}\sqrt{\frac{D_{o}}{\pi}} = D_o \sqrt{\frac{2 t_c}{\pi}},
\end{equation}
where $l_c$ is defined as the average inter-particle separation and $t_c = l_c^2/2 D_0$ is the average time between collisions.  On time scales $t$ much less than the collision time $t_c$, one expects  ordinary Fickian diffusion, whereas for time scales $t \gg t_c$ one expects to observe  the asymptotic SFD diffusion of eq.~(\ref{eq:sfdmsd}).

An obvious question raised by the experiments on SFD is what happens as the confinement becomes less severe.  At some point the system should cross over to ordinary Fickian diffusion at long time-scales.  This problem was first studied by coupling two lattice gas models so that particles could jump between chains~\cite{Kutn84}.  Allowing the jumps produced a crossover in the mean-square displacement from  subdiffusive $\sqrt{t}$ scaling to diffusive $\lb x^2 \rb ~\sim t$ asymptotic scaling at long times.  Difficulties in interpreting measurements of SFD transport in zeolites also inspired simulations that exhibited a similar long-time crossover to Fickian diffusion~\cite{Hahn98,Tepp99}.

As shown by Mon and Percus~\cite{Mon02}  it is convenient to analyze  this crossover  in terms of a Markov process with a hopping time $t_{hop}$ that measures the average time for two particles to pass one another.  If $t_{hop} \gg t_c$, so that the system has reached the SFD regime between hops, then  the average mean-square displacement of the particles scales as $l_{hop}^2 \sim F t^\frac12_{hop}$ between hops.  Within this picture a particle makes on average $t/t_{hop}$ hops in a time $t$ so that its mean-square displacement scales as
\begin{equation}\label{eq:hop}
\lb x^2 \rb \sim l_{hop}^2\left( \frac{t}{t_{hop}}\right) \sim D_{hop} \, t
\end{equation}
which defines an effective Fickian diffusion coefficient  of the form
\begin{equation}\label{eq:Dhop}
D_{hop} \sim l^2_{hop}/t_{hop} \approx F t_{hop}^{-\frac12}.
\end{equation}
The larger $t_{hop}$, the smaller $D_{hop}$, since particles passing events become more rare.  Nevertheless, on time-scales $t \gg t_{hop}$ the final asymptotic scaling will still be Fickian. On the other hand, if $t_{hop}$ decreases to the point where it is of the order of the collision time $t_c$, then the SFD picture is expected to break  down and  $D_{hop}$ will approach the self-diffusion coefficient $D_0$.  So for the hopping picture to be useful, we require  $t_{hop} \gg t_c$.   

The hopping time $t_{hop}$ has been studied in some depth for the case of two discs diffusing between hard walls a distance $L$ apart. Mon {\em et al}.~\cite{Mon02,Bowl04,Mon06,Mon08b} found a scaling of the form
\begin{equation}\label{eq:thopscaling}
t_{hop} \sim (L - 2 \sigma)^{-\nu} = \delta^{-\nu}
\end{equation}
as the pore width $L$ approaches the limit where two particles can no longer pass. An exponent $\nu = 2$ was found both for Molecular dynamics~\cite{Bowl04} and Brownian dynamics (BD)~\cite{Mon08b}, but the interpretation of the simulations is subtle~\cite{Mon08a}.  This exponent agrees with direct calculations of the diffusion equation~\cite{Mon06}, and a simpler transition state theory (TST)~\cite{Bowl04,Kali07}, but not with the effective one-dimensional Fick-Jacobs equation~\cite{Bowl04,Kali07,Kali08}, which predicts that $\nu = \frac32$.

Given that hydrodynamic interactions (HI) can have subtle effects on the dynamics of colloidal particles~\cite{Russ89,Dhon96}, 
one might expect there to be differences between SFD behaviour measured for colloidal suspensions, and the SFD behaviour of particles where hydrodynamics is not important.  A careful theoretical study by Kollmann~\cite{Koll03} predicted that the HI  do not change the asymptotic scaling of Eq.~(\ref{eq:sfdmsd}), and moreover that the mobility $F$ can be connected to the short-time collective diffusion coefficient of the system.   Indeed, the experiments  with colloidal particles cited above exhibit SFD sub-diffusive behaviour, as expected. 
Nevertheless, it would be interesting to see if other aspects of SFD behaviour are sensitive to HI.
Moreover, the subtle crossover from SFD to Fickian long-time diffusion for narrow pipes where particles can hop past each other, but where $t_{hop} \gg t_c$, could well be affected by HI.

To study these questions, we employ a hybrid MD and stochastic rotations dynamics (SRD) computer simulation method that has been shown to accurately reproduce Brownian and hydrodynamic behaviour for colloidal suspensions~\cite{Male00,Padd06}.  SRD, also known as multi-particle collision dynamics, was first described by Malevanets and Kapral~\cite{Male99}.
SRD has been applied to a wide number of different systems, including fluid vesicles in shear
flow~\cite{Nogu04}, clay-like colloids~\cite{Hech05}, sedimentation of
colloids~\cite{Padd04,Wyso08}, colloidal rods in shear
flow~\cite{Ripo08}, knots in viral DNA~\cite{Matt09} and many other
examples~\cite{Kapr08}.  We have recently used this method to  study the role of confinement on two-dimensional diffusion~\cite{Sane09a} and the role of finite sized particles on Taylor dispersion~\cite{Sane09b}.

 To capture the correct asymptotic trends for the mean-square displacement, very long simulation runs are needed, which explains why so many of the simulations in the literature have been on two-dimensional models, which are generally faster to simulate than three-dimensional systems.  In this paper we study strongly repulsive colloidal discs in confinement, using both the hybrid MD-SRD method as well as simpler BD to compare what happens when HI are ignored.   We find, in agreement with experiments on colloidal suspensions~\cite{Lin05}, that the simple hard-rod model for the SFD mobility~(\ref{eq:mobhr}) provides a good fit  both for BD and for the simulations that include HI.  

When the pipe diameter is such that $\delta = L-2 \sigma > 0$, so that the mutual passage constraint is lifted, we observe a crossover from SFD sub-diffusion to Fickian diffusion at longer times $t \gtrsim t_{hop}$.  We measure the distribution of hopping times $t_{hop}$, showing that they follow Poissonian statistics for simulations with and without hydrodynamic interactions.  For both models we also find a scaling consistent with $D_{hop} \propto 1/\sqrt{t_{hop}}$.   For the BD simulations, we find that $t_{hop} \sim \delta^{-2}$ for small $\delta/\sigma$, but $t_{hop}$ depends more strongly on $1/\delta$ when hydrodynamic interactions are included.  We attribute this to repulsive lubrication forces, induced when particles approach each other other or the walls of the container.

The paper is organised as follows:  In section II, we describe the model, our hybrid MD-SRD method and our implementation of BD.  In section III  we study the SFD behaviour of colloids confined to one-dimensional flows for different packing fractions $\eta$. Section IV considers the case when the mutual passage constraint is lifted, so that at long times Fickian diffusion is recovered.   Finally, in section V, we discuss the main conclusions of this study.

\section{Model and Simulation Methods}\label{sec_method}
\subsection{Model}

For the colloid-colloid interactions, we use a very strongly repulsive potential of the form:
\begin{displaymath}
\varphi_{cc}(r) = \left\{
\begin{array}{ll}
4\epsilon_{cc}\left( \left(\frac{\sigma}{r}\right)^{48}-
\left(\frac{\sigma}{r}\right)^{24} +\frac{1}{4}\right) & (r\leq 2^{1/24}\sigma)\\
0 & (r\geq 2^{1/24}\sigma)
\end{array} \right.
\end{displaymath}
where $\sigma$ is the colloid-colloid diameter.  The colloid-wall interaction is taken to be strongly repulsive as well, and a stick-boundary condition is applied to the colloids, as well as to the fluid particle-wall interaction in the SRD simulations.  We use periodic boundary conditions in the direction parallel to the pipe walls.

\subsection{Hybrid MD and SRD simulation method}

To describe the hydrodynamic behaviour of colloids induced by a background fluid of much smaller constituents, some form of coarse-graining is required. The hydrodynamics can be described by the
Navier Stokes equations that coarse-grain the fluid within a continuum description.  The downside of going directly through this route is that every time the colloids move, the boundary conditions on
the differential equations change, making them computationally expensive to solve.

An alternative to computing a direct solution of the Navier Stokes equations is to use particle based techniques that exploit the fact that only a few conditions, such as (local) energy and momentum conservation, need to be satisfied to allow the correct (thermo) hydrodynamics to emerge in the continuum limit. Simple particle collision rules, easily amenable to efficient computer simulation, can therefore be used.  Boundary conditions (such as those imposed by colloids in suspension) are easily implemented as external fields.

In this paper we implement the SRD method first derived by Malevanets and Kapral~\cite{Male99}.  An SRD fluid is modelled by $N$ point particles of mass $m$, with 
positions ${\bf{r}}_{i}$ and velocities ${\bf{v}}_{i}$. The coarse graining procedure consists of two steps, streaming and collision. During the streaming step, the positions of the fluid particles are updated via
\begin{equation}
{\bf{r}}_{i}(t+\delta t_s) = {\bf{r}}_{i}(t) + {\bf{v}}_{i}(t)\delta t_s.
\end{equation} 
In the collision step, the particles are split up into cells with sides of length $a_0$, and their velocities are rotated around an angle $\alpha$ with respect to the cell centre of mass velocity,
\begin{equation}
{\bf{v}}_{i}(t+\delta t_s) = {\bf{v}}_{c.m,i}(t) +
\mathcal{R}_i(\alpha)\left[{\bf{v}}_{i}(t)-{\bf{v}}_{c.m,i}(t)\right] 
\end{equation}
where ${\bf{v}}_{c.m,i}=\sum^{i,t}_{j}(m{\bf{v}}_{j})/\sum_{j}m$ is the centre of mass velocity of the particles in each cell, $\mathcal{R}_i(\alpha)$ is the rotational matrix and $\delta t_s$ is the interval between collisions. The purpose of this collision step is to transfer momentum between the fluid particles while conserving the energy and momentum of each cell.

The fluid particles only interact with one another through the collision procedure. Direct interactions between the solvent particles are not taken into account, so that the algorithm scales as
$\cal{O}$$(N)$ with particle number.  The carefully constructed rotation procedure can be viewed as a coarse-graining of particle collisions over space {\em and} time.  Mass, energy and momentum are
conserved locally, so that on large enough length-scales the correct Navier Stokes hydrodynamics naturally emerges~\cite{Male99}.  It is important to remember that for all these particle based methods, the  particles should not be viewed as composite supramolecular fluid units, but rather as coarse-grained Navier Stokes solvers (with noise in the case of SRD)~\cite{Padd06}. 
Another advantage of SRD is  that transport coefficients
have been analytically calculated~\cite{Ihle03,Kiku03,Pool04}, greatly
facilitating its use. 

The SRD fluid particles  can easily be coupled to a solute as
first shown by Malevanets and Kapral~\cite{Male00}, and studied in detail in a recent paper~\cite{Padd06}.  To simulate the behaviour of 
colloids of mass $M$,  we use the colloid-colloid interaction defined above, while the solvent particles interact with the colloids via an interaction of the form:
\begin{displaymath}
\varphi_{cs}(r) = \left\{
\begin{array}{ll}
4\epsilon_{cs}\left( \left(\frac{\sigma_{cs}}{r}\right)^{12}-
\left(\frac{\sigma_{cs}}{r}\right)^{6} +\frac{1}{4}\right) & (r\leq 2^{1/6}\sigma_{cs})\\
0 & (r\geq 2^{1/6}\sigma_{cs})
\end{array} \right.
\end{displaymath}
where $\sigma_{cs}$ is the 
colloid-solvent collision diameter.  We propagate the ensuing equations of
motion with a Velocity Verlet algorithm~\cite{Allen} using a molecular
dynamic time step $\Delta t$
\begin{eqnarray} R_{i}(t+\Delta t) &=&
  R_{i}(t) + V_{i}(t)\Delta t + \frac{F_{i}(t)}{2M}\Delta t^{2}\\
  V_{i}(t+\Delta t) &=& V_{i}(t) + \frac{F_{i}(t)+F_{i}(t+\Delta
    t)}{2M}\Delta t,
\end{eqnarray}
where $R_{i}$ and $V_{i}$ are the position and velocity of the colloid, and
$F_{i}$ is the total force exerted on the colloid.  Coupling the colloids in this
way leads to slip boundary conditions.  Stick boundary conditions can also be
implemented~\cite{Padd05}, but for qualitative behaviour, we don't expect there
to be important differences.  In parallel the velocities and positions of the
SRD particles are streamed in the external potential given by the colloids and
the external walls and updated with the SRD rotation-collision step every
time-step $\delta t_s$.

The larger the ratio $\sigma/a_0$, the more accurately the hydrodynamic
flow fields will be reproduced. In ref~\cite{Padd06}, it is shown that using
$\sigma/a_0 = 4.3$, and $\sigma_{cs} = 2 a_0$ reproduces the flow fields
with small relative errors for a single sphere in a 3d flow. Because we are interested in processes where the colloids can just barely pass each other, we chose a finer grid  ($\sigma/a_0=8.6$). This should in particular enhance the resolution of lubrication forces~\cite{Padd06}.  Other parameter choices taken from Refs~\cite{Padd06,Sane09a} include
$\epsilon_{cc}=\epsilon_{cs} = 2.5 k_BT$ for the colloids, an SRD particle density of $\gamma = 5$ particles per $a_0^2$ and a rotation angle of 
$\alpha = \frac12 \pi$.  The time-steps for the MD and SRD step are set by different physics~\cite{Padd06}, and we chose $\Delta t = 0.025 t_0$ and
$\delta t_s = 0.1 t_0$ for SRD, where $t_0 = a_0 \sqrt{\frac{m}{k_B T}}$ is the
unit of time in our simulations.

\subsection{Brownian Dynamics simulation method}

The particle motion for the colloids can also be solved using BD.  The positions are updated via the equation of motion:
\begin{equation}
r(t+\Delta t_{BD}) = r(t) + \frac{\Delta t_{BD}}{m\xi}F(t) +  \delta r^{G},
\end{equation}
where $F(t)$ is the force acting on the
colloids, which arises from the colloid-colloid interaction, as well as the colloid-wall interactions,  and each component $\delta r^{G}$ is chosen from a Gaussian
distribution with zero mean and variance $\lb(\delta r^{G})^{2} \rb =
2D_{o}\Delta t$ ~\cite{Allen}, where $D_0$ is the single particle diffusion coefficient. 
With BD, the particles execute a random walk with a physical time-scale 
\begin{equation}\label{eq:taud}
t_D = \frac{\sigma^2}{2 D_0}
\end{equation}
called the diffusion time that measures how long it takes the particle to diffuse over its diameter.
 The BD time-step was chosen to be $\Delta t_{BD} = 2\times 10^{-5}t_D$, slightly smaller than the value used, for example, in a careful study by Lodge and Heyes~\cite{Lodge}.  We use this conservative measure of the time-step because there are indications that scaling of the hopping time of colloidal particles simulated with BD is sensitive to the time-step~\cite{Mon08a}; we checked that our time-step is small enough.
In contrast to the SRD method, which also exhibits diffusive motion, the BD  method does not include HI.  

\begin{figure}
\begin{center}
\includegraphics[angle=-90,width=0.5\textwidth]{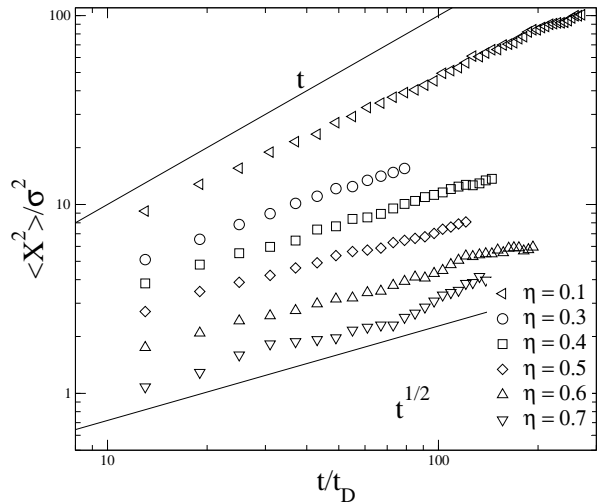}
\caption{\label{Fig:SFDBDENS} Log-log plot of the mean square displacement of
colloids calculated with BD for a pipe of width $L=1.4 \sigma$.  
 Results are shown for simulations with increasing line packing fractions
$\eta = 0.1,0.3,0.4,0.5,0.6,0.7$.  The straight lines have slope $1$ and $1/2$
respectively and serve as a guide to the eye.  At long times the colloids show clear signatures of SFD sub-diffusive scaling. }
\end{center}
\end{figure}

\begin{figure}
\begin{center}
\includegraphics[angle=-90, width=0.5\textwidth]{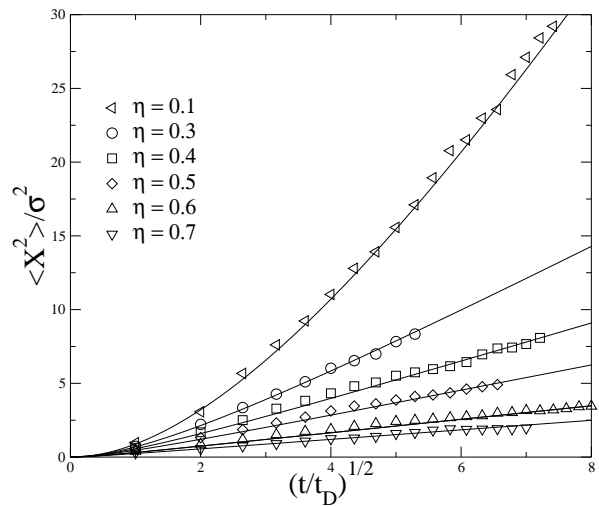}
\caption{\label{Fig:SFDBD} Linear plot of the mean square displacement of
colloids as a function of $\sqrt{t/t_D}$, calculated with BD for a pipe of width $L=1.4 \sigma$.  
 Results are shown for simulations with increasing line packing fractions
$\eta = 0.1,0.3,0.4,0.5,0.6,0.7$. The solid lines are a fit from Eq.~(\ref{eq:ansatz}). Since $D_0$ is fixed by the BD simulation parameters  the only fit parameter is the SFD mobility $F$. }
\end{center}
\end{figure}

\begin{figure}
\begin{center}
\includegraphics[angle=-90, width=0.5\textwidth]{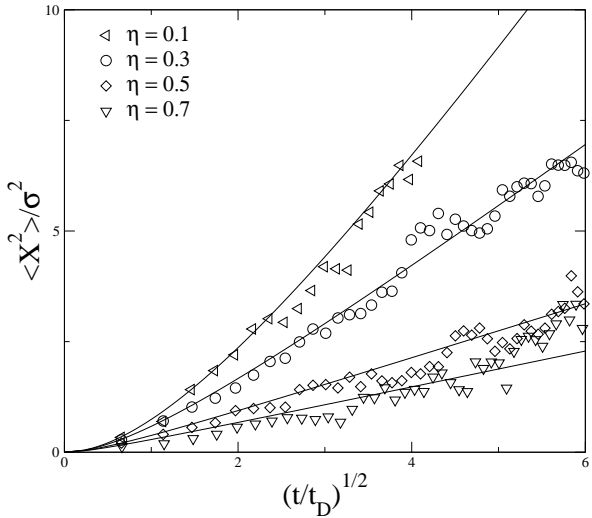}
\caption{\label{Fig:SFDSRD} Linear plot of the mean square displacement of
colloids as a function of $\sqrt{t/t_D}$, calculated with SRD for a pipe of width $L=1.4 \sigma$.  In contrast to the BD simulations, these include HI. The solid lines are the fit to  Eq.~(\ref{eq:ansatz}).  }
\end{center}
\end{figure}

\begin{figure}
\begin{center}
\includegraphics[angle=-90, width=0.5\textwidth]{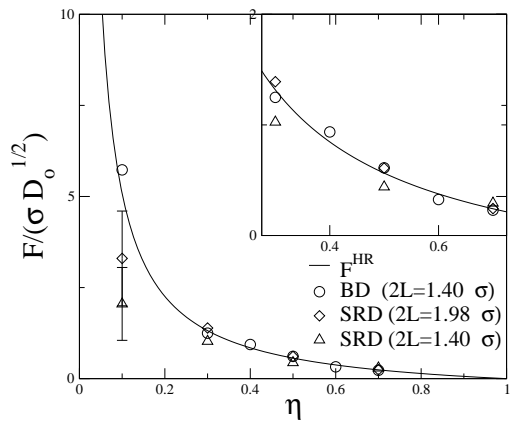}
\caption{\label{Fig:FSRDBD2} Comparison of the SFD particle mobility $F$, extracted from BD and SRD simulations, to the $F^{HR}$ for a hard-rod fluid from Eq.~(\ref{eq:mobhr}).   Inset: data at higher packing fractions. The results suggest that, within simulation errors, the inclusion of HI interactions does not strongly affect the 1D mobility $F$. }
\end{center}
\end{figure}

\section{Single file diffusion}

We first investigate the SFD behaviour of colloidal discs in pipes narrow enough that they cannot pass each other.   At times less than the collision time $t_c$  we expect the particles to behave in a diffusive manner, and for times $t \gg t_c$ we expect SFD sub-diffusion.   Lin {\em et al.}~\cite{Lin05} showed that the following ansatz:
\begin{equation}
\frac{1}{\lb x^{2}\rb} = \frac{1}{2D_{o}t} + \frac{1}{2Ft^{1/2}}
\end{equation}
provides a good approximation to the crossover from Fickian to SFD sub-diffusion.
Solving for the mean-square displacement gives:
\begin{equation}\label{eq:ansatz}
\lb x^{2}\rb = \frac{2D_{o}t}{1+(D_{o}/F)t^{1/2}} = \frac{2D_{o}t}{1+(t/t_{x})^{1/2}},
\end{equation} 
where we have defined the crossover time 
\begin{equation}
t_{x} = (F/D_{o})^{2},
\end{equation} 
a measure of the time needed for the system to transition from Fickian diffusion to the asymptotic SFD regime.
For particles that behave like hard rods with a mobility given by Eq.~(\ref{eq:mobhr}), this crossover time  scales as:
\begin{equation}
t_x = t_{D} \frac{2}{\pi}\left(\frac{1-\eta}{\eta}\right)^2,
\end{equation}
and is connected to the collision time by $t_x = 2 t_c/\pi$.
That the two times are essentially the same is not surprising, given that the
$\left\langle x^2 \right\rangle \sim \sqrt{t}$ behaviour is generated by the collisions
with neighbours and the long-ranged correlations these induce.
 The higher the one-dimensional packing fraction $\eta$, the lower $t_x$, and so the more rapidly the system should transition to asymptotic SFD sub-diffusion.
For very small packing fractions the crossover time scales with $t_x/t_D \sim 1/\eta^2$, and so very long simulations are necessary to observe asymptotic SFD scaling.  For example, for $\eta = 0.5$, $t_x \approx 0.63 t_D$, for $\eta=0.1$, $t_x \approx 51.5 t_D$ and for $\eta =0.01$, $t_x \approx 6240 t_D$.

 Simulations were carried out with BD for  $N=200$ colloidal particles
diffusing between parallel plates a distance $L = 1.40 \sigma$
apart. The one-dimensional packing fraction was varied in the range
$\eta=\frac{N\sigma}{L_p}=0.1-0.7$ and, after equilibration, simulations were run for total times ranging between   $100t_D$ up to $300t_D$ to gather data on the mean-square displacement.

In Fig.~\ref{Fig:SFDBDENS}, we plot the mean square displacement of particles
undergoing single file diffusion at various packing fractions.  With the exception of the data for $\eta = 0.1$, the crossover time $t_x \ll 10 t_D$ so that SFD $\lb x^2 \rb \sim \sqrt{t}$ scaling is expected for the entire range of data plotted.  The measurements are consistent with this scaling.  The same data is plotted on a linear plot in Fig.~\ref{Fig:SFDBD}, where it is compared to the ansatz of Eq.~(\ref{eq:ansatz}).     Since $D_0$ is given by the simulation parameters, there is only one fit parameter, namely the SFD mobility $F$.   The plots demonstrate the  $\sqrt{t}$ scaling for the mean square displacement v.s.\ time.   

Simulations were also carried out  with SRD.  The explicit inclusion of a background fluid induces  HI between the particles, and between the particles and the walls.  The combination of strong confinement and HI  has an important qualitative effect on the velocity autocorrelation functions~\cite{Pago01,Sane09a} and on the Fickian diffusion coefficient~\cite{Bung73,Boc} of individual particles, especially in two dimensions~\cite{Saff76,Sane09a}, where HI are particularly important.  But once the system is confined so strongly that particles can no longer pass, the long-ranged HI interactions can be strongly screened~\cite{Cui02}.   That raises the question:  what is the effect of including HI on the SFD mobility $F$?   The SRD method allows us to address these issues.

We performed simulation runs for  $N=200$ colloidal particles in a SRD fluid
between parallel plates a distance $L= 1.40 \sigma$ and $L=1.98 \sigma$ apart.  The longitudinal packing
fractions were the same as for the BD simulations.  After equilibration, simulation data was gathered for runs of approximately  $10-30t_D$ in length.  Because the SRD method includes many solvent particles (${\cal{O} }(10^5)$ here), it is computationally more expensive than the BD simulations, and so the runs are shorter than those performed for BD.  The mean-square displacements  for the $L = 1.4 \sigma$ case are shown in 
in Fig.~\ref{Fig:SFDSRD}, and  $\lb x^2 \rb \sim \sqrt{t}$ behaviour is still observed.   In this case a fit of the data to Eq.~(\ref{eq:ansatz}) was used to extract  both $D_0$ and $F$.  The values of $D_0$ are consistent with those of ref.~\cite{Sane09a}.

The values of the fits to $F$ for both BD and SRD are shown in Fig.~\ref{Fig:FSRDBD2} as a function of the packing fraction $\eta$.    These are compared to the hard rod $F^{HR}$ of Eq.~(\ref{eq:mobhr}).  At all but the lowest packing fractions, good agreement is found, suggesting that the inclusion of HI does not significantly affect the value of $F/D_0$.  Moreover, in contrast to what is found in two dimensions where HI interactions have an important impact on the value of $D_0$~\cite{Saff76,Sane09a}, here the measured value of the normalised SFD mobility  $F/(\sqrt{D_{0}}\sigma)$ does not seem to depend significantly on the confinement, since differences between what is measured at $L = 1.4 \sigma$ and at $L=1.98 \sigma$ appear to be small, even though $D_0$ changes considerably~\cite{Sane09a}.   Note that $F$ itself does scale with $\sqrt{D_0}$, so the changes in $D_0$ due to confinement translate into changes in $F$, but there are  no further effects of the HI.      Finally, we should point out that some further changes in $F$ are expected when $L$ moves away from the pure hard rod limit $L=\sigma$, but these are expected to be small~\cite{Mon07}.

 It is only for the lowest packing fraction $\eta = 0.1$ that we observe significant differences with $F^{HR}$ for the SRD simulations.  Interestingly, this deviation is very similar to that found by Lin {\em et al.}~\cite{Lin05} in their experiments.    Unfortunately, at this lower packing fraction very long simulations are needed to reach the asymptotic SFD regime, and so we don't believe that enough data has been gathered here to reliably extract $F$.  The error bars shown in the graph are from a standard non-linear fitting procedure, but it is very likely that the real error is higher.   A similar conclusion about the errors was made by Lin {\em et al.}~\cite{Lin05} for their data, which also extended out to $t \approx t_x$ for the lowest packing fraction.  To resolve this question, the simulations and the experiments should be performed for  at least an order of magnitude longer time than they have been.

\begin{figure}
\begin{center}
\includegraphics[width=0.25\textwidth]{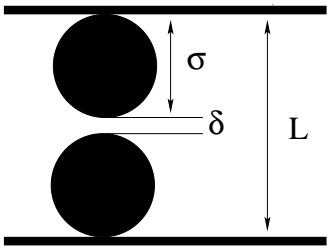} 
\caption{\label{Fig:delta}Schematic depiction of the clearance $\delta$
available to particles attempting to switch places with one another. Particles
can hop to pass their neighbours if $\delta > 0$.}
\end{center}
\end{figure}

\begin{figure*}
\begin{center}
\includegraphics[angle=-90, width=0.75\textwidth]{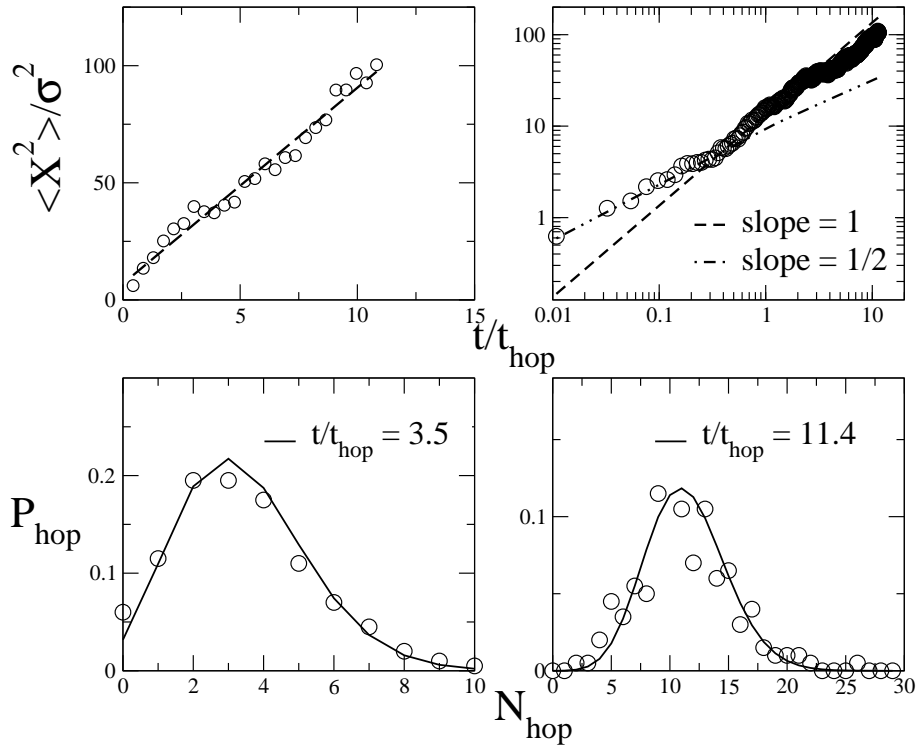}
\caption{\label{Fig:hop1}Crossover from SFD to Fickian diffusion.  The top two plots
show the mean square displacement of the particles calculated with a BD
simulation at $\eta = 0.5$ for $\delta_c = 0.093$.    The
mean hopping time was measured  to be $t_{hop}\approx 86t_D$.  On timescales of the order of the hopping time, we observe a crossover to Fickian diffusion
$\lb x^2 \rb \sim t$  with an effective diffusion coefficient $D_{hop}\approx 0.095 D_0$.  This crossover is especially clear in the log plot on the top right. .  The
bottom two plots show the probability $P_{hop}$ that a particle has performed $N$ hops in the time intervals  
intervals, $t\approx 3.45 t_{hop} $ and $t\approx 11.43 t_{hop}$ respectively.}
\end{center}
\end{figure*}

\begin{figure*}
\begin{center}
\includegraphics[angle=-90, width=0.75\textwidth]{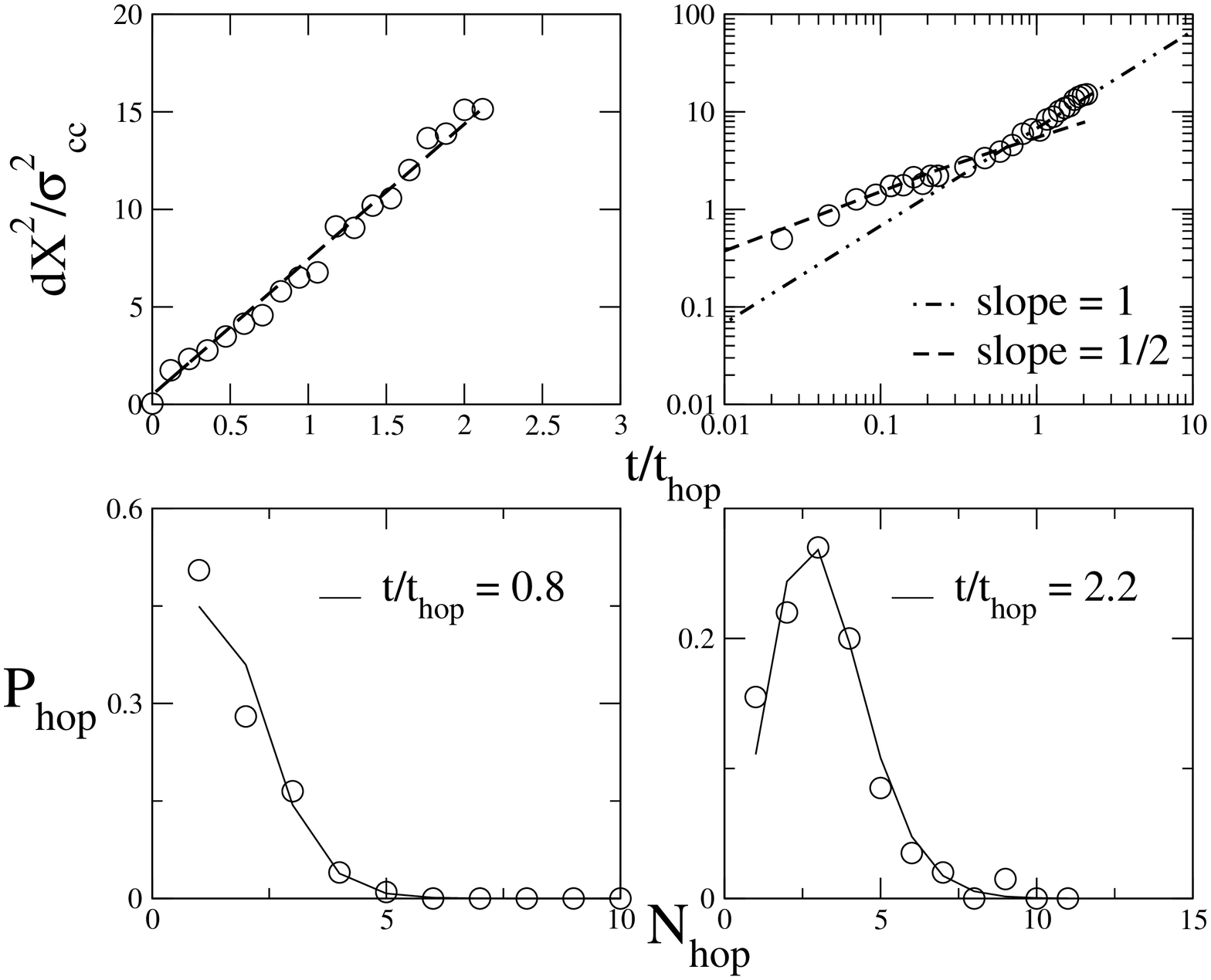}
\caption{\label{Fig:hop1srd}Crossover from SFD to Fickian diffusion for simulations that  include hydrodynamic interactions. The top two plots
show the mean square displacement of the particles calculated with an SRD 
simulation at $\eta = 0.5$ for $\delta_c = 0.093$.  The
mean hopping time was measured  to be $t_{hop}\approx 79t_D$.  The
bottom two plots show the probability $P_{hop}$  that a particle has performed $N$ hops in the time intervals  
intervals, $t\approx 0.8 t_{hop} $$ (6.3 t_D)$ and $t\approx 2.2 t_{hop}$$ (17 t_D)$ respectively.
 The straight line
shows the predicted Poissonian distribution~(\ref{eq:poisson}) and the circles are the results
from the simulation. }
\end{center}
\end{figure*}

\section{Crossover from SFD to  Fickian diffusion}

Having worked out some properties of SFD when particles cannot pass each other, we now focus on the dynamics that
emerge when the single file constraint is lifted.  
Fig.~\ref{Fig:delta} illustrates how  the parameter $\delta_c = \delta/\sigma = L/\sigma - 2$ describes the maximal  distance between the particles when they can pass each other. In our case we don't quite have hard sphere particles, but the $1/r^{48}$ repulsion is hard enough that the differences are expected to be very small. So we will still use the parameter $\delta_c$ to denote how easy it is for particles to pass.

\subsection{Scaling of the mean-square displacement}

 As shown in the introduction, it is very useful to consider the hopping time $t_{hop}$ that it takes a particle to switch order with one of its neighbours.  
From Eq.~(\ref{eq:Dhop}) it follows that the effective Fickian diffusion coefficient is expected to scale as $D_{hop} \sim F t_{hop}^{-\frac12}.$

To test this concept we performed a number of simulations at different values of $\delta_c$.  As an illustrative example, we show the mean-square displacement  of $N=200$ Brownian particles at $\eta = 0.5$ and  $\delta_c = 0.093$ in Fig.~\ref{Fig:hop1}.   In this case the hopping time  $t_{hop} \approx 86 t_D \gg t_x \approx 0.12 t_D$, so that the crossover from the initial Fickian diffusion with a diffusion coefficient of $D_0$  to SFD  at $t \approx t_x$ is at much shorter times than an eventual longer time crossover back to diffusive motion due to the hopping.  We simulate for up to $t=1040 t_D$ so that  there are on average about $11.4$ hopping processes per particle.  
At times on the order of $t_{hop}$ we observe a  crossover from $\lb x^2 \rb \sim \sqrt{t}$ to the asymptotic linear Fickian diffusive scaling $\lb x^2 \rb \sim 2 D_{hop} t$ with an effective diffusion coefficient $D_{hop} \approx 0.095 D_0$.    Although in this case a hopping event occurs only about once in every thousand collisions, the net effect is to generate a linear diffusion that at longer times leads to a significantly larger mean square displacement than if the particles couldn't pass.

Fig.~\ref{Fig:hop1srd} shows a similar set of simulations at $\eta = 0.5$, but now with the  SRD methods that includes  HI.  Although the runs are shorter (SRD is computationally more expensive), they nevertheless show the same trends as the BD simulations, with a crossover to Fickian diffusive motion at $t \gtrsim t_{hop}$
with an effective diffusion coefficient $D_{hop}\approx 0.087 D_0$.  This scaling behaviour is especially clear in the log plot on the top right.  For this set of parameters, the SRD and the BD give similar results for $t_{hop}$ and $D_{hop}$ within the expected errors of the simulations.
 
 \subsection{Poissonian statistics of hopping events}

 The order of the particles was tracked and the number of times that a particle switched places with its neighbour was recorded.  This was done on an interval large enough to avoid double counting short switches during the hopping process.  If the hopping events are independent, then  the probability that a particle makes $N$ hops is proportional to the time interval over which a measurement is made and should follow Poissonian statistics.  The probability that such an event will occur $n$ times in a given time interval
can be expressed as
\begin{equation}\label{eq:poisson}
P(n;\lambda) = \frac{\lambda^{n}e^{-\lambda}}{n!}.
\end{equation}
where $\lambda = t/t_{hop}$ in our case.
The bottom two plots of Figs.~\ref{Fig:hop1} and ~\ref{Fig:hop1srd} show the distributions  at two different values of $\lambda$. The
solid lines denote the distributions calculated from the Poisson distribution~(\ref{eq:poisson}),
and the circles denote those obtained from the simulation. Within the simulation error, these  show good agreement for both the BD and the SRD simulations, suggesting that the hopping events are indeed independent and not correlated with each other.

\begin{figure}
\begin{center}
\includegraphics[angle=-90,width=0.5\textwidth]{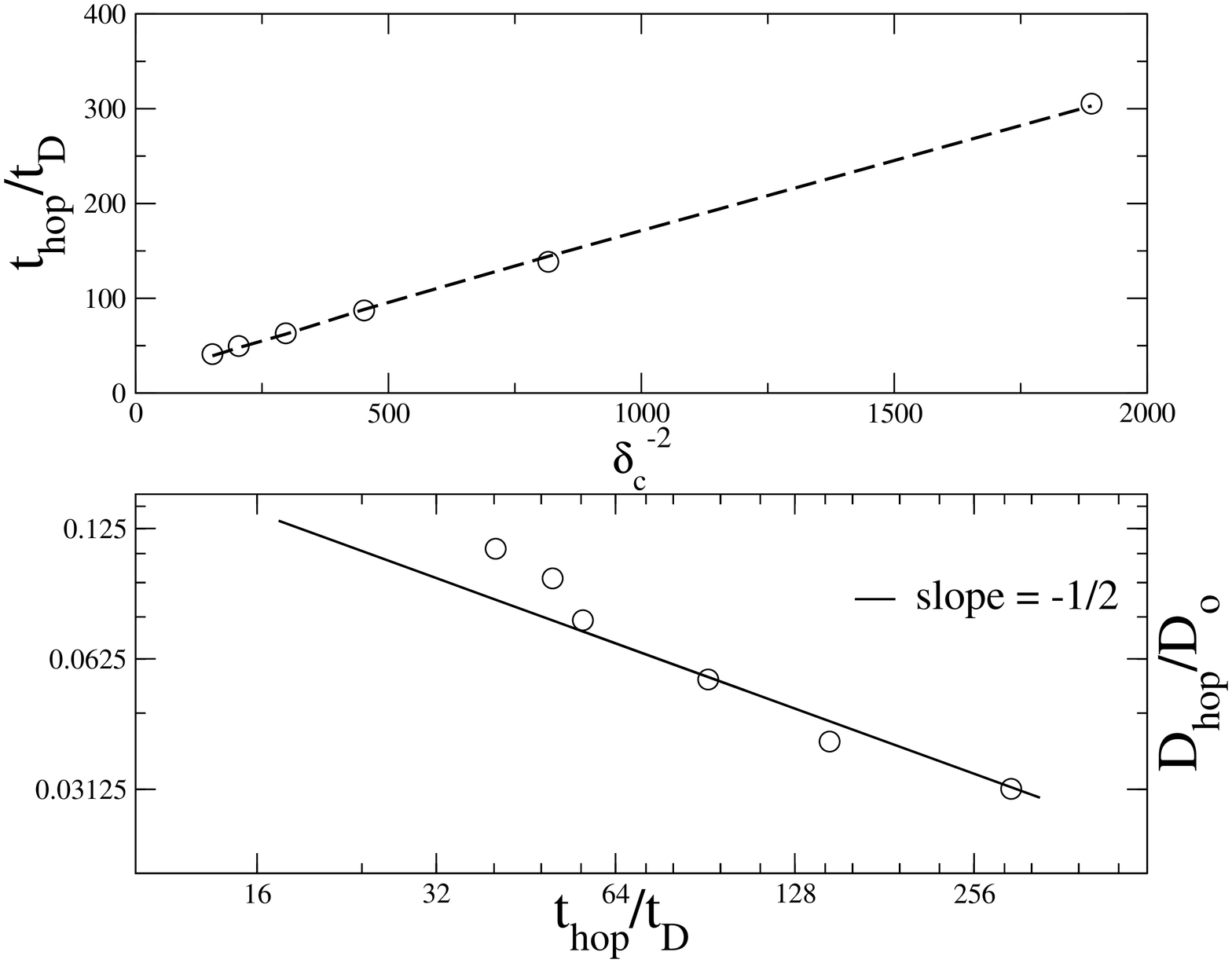}
\caption{\label{Fig:thopdhop}  BD simulations  of the hopping time $t_{hop}$ and the effective diffusion coefficient $D_{hop}$  for packing fraction $\eta = 0.7$ and separations $\delta_c = 0.023,0.035,0.047,0.058,0.070$, and $0.081$. Top: Average hopping time  plotted versus $1/\delta_c^2$ 
The straight line is a fit to the data.  Bottom: Effective diffusion coefficient $D_{hop}$ v.s. $t_{hop}$
The straight line has a slope $-1/2$ and serves as a guide to the eye. }
\end{center}
\end{figure}

\begin{figure}
\begin{center}
\includegraphics[angle=-90, width=0.5\textwidth]{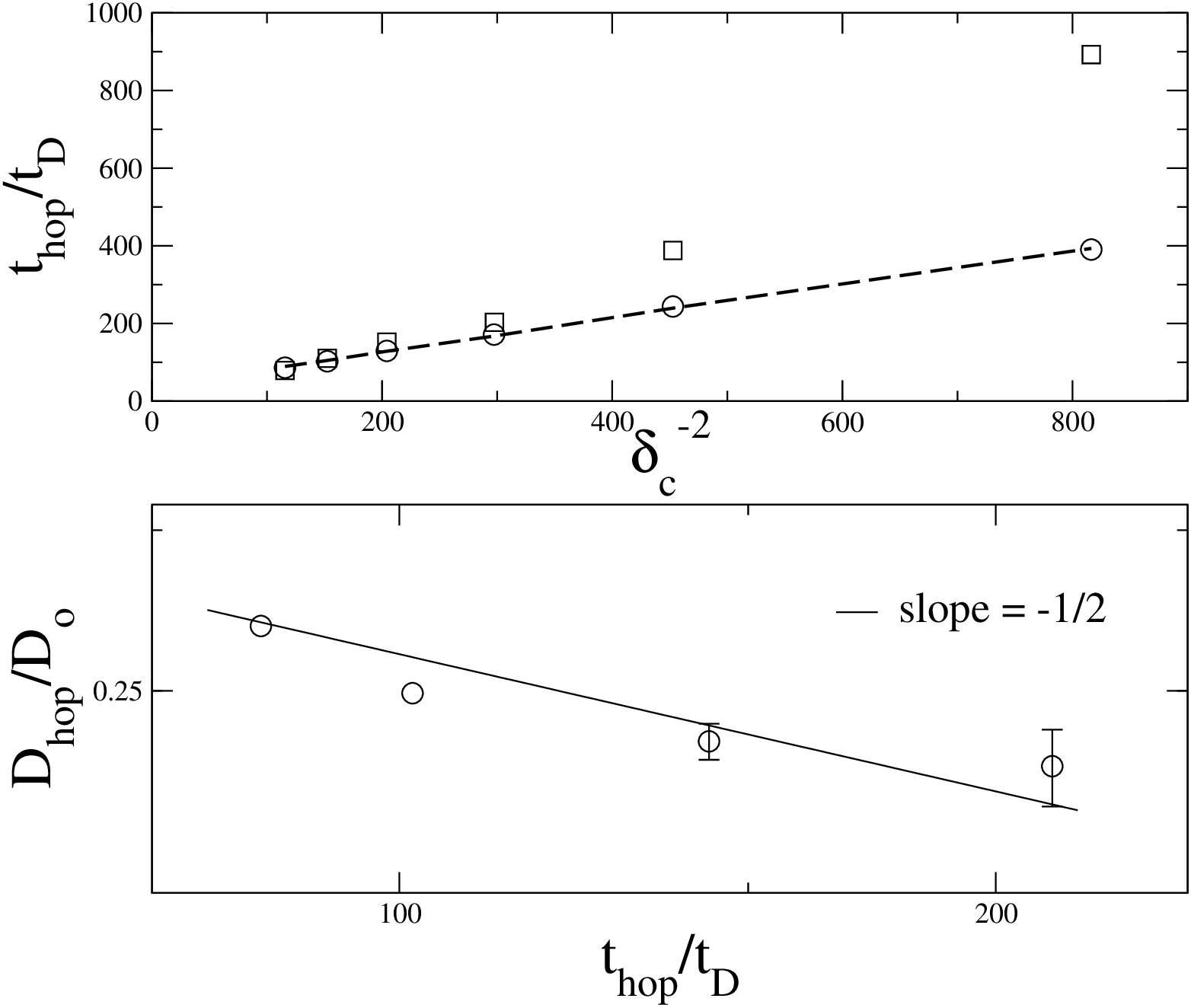}
\caption{\label{Fig:thopbdsrd2}
 Top graph: SRD and BD simulations  of the hopping time $t_{hop}$ for packing fraction $\eta = 0.5$ and separations $\delta_c = 0.035,0.047,0.058,0.070,0.081$, and $0.093$.   For larger $\delta_c$ both simulation methods give the same value of $t_{hop}$, but at smaller separations the simulations with HI have larger $t_{hop}$'s.
 Bottom: Effective diffusion coefficient $D_{hop}$ v.s. $t_{hop}$ from the SRD simulations.  The straight line has a slope $-1/2$ and serves as a guide to the eye; The data is s consistent with $D_{hop}\sim t_{hop}^{-1/2}$.   
 }
\end{center}
\end{figure}

\subsection{Scaling of $t_{hop}$ and $D_{hop}$ with plate separation}

Further simulations were performed for 1000 Brownian particles diffusing in pipes of various sizes.
In  Fig.~\ref{Fig:thopdhop} we plot the hopping times $t_{hop}$ and effective diffusion coefficients $D_{hop}$ for BD simulations at $\eta = 0.7$.  The top panel shows how hopping time varies with the distance $\delta_c$ and is consistent with $t_{hop} \propto \delta_c^{-2}$.  
This exponent agrees with direct calculations of the diffusion equation~\cite{Mon06}, and a simpler transition state theory~\cite{Bowl04,Kali07}, but not with the effective one-dimensional Fick-Jacobs equation~\cite{Bowl04,Kali07,Kali08}.   
Nevertheless, we don't consider our simulations to be on a large enough range of $\delta_c$ to conclusively determine the scaling behaviour.  Such simulations become increasingly difficult for smaller $\delta_c$ because the hopping events become more rare.  However, recent Monte Carlo simulations by Mon~\cite{Mon08a} for just two discs  are consistent with transition state theory over a sufficiently larger range of $\delta_c$ to confirm the scaling law for that special geometry.  His results suggest that if full simulations of many discs were performed over a wider range of $\delta_c$ they would confirm the scaling law we observe over a limited range of $\delta_c$.

In the bottom panel in Fig.~\ref{Fig:thopdhop}, we plot the effective diffusion coefficient $D_{hop}$
versus $t_{hop}$. The solid line has the slope $-1/2$ which is consistent with the scaling of  Eq.~(\ref{eq:Dhop}), as postulated by Mon and Percus~\cite{Mon02}.  Again, the range of $\delta_c$'s is not large enough to firmly fix the scaling behaviour, but it is at least consistent with $D \propto t^{-\frac12}$ scaling, and inconsistent with an earlier prediction of $t^{-1}$ scaling~\cite{Hahn98}.
 As the pipes become wider and $t_{hop}$ becomes smaller, we expect $D_{hop}$ to eventually increase to $D_0$, the self-diffusion coefficient. 
 Thus  at larger $\delta_c$ this simple scaling law should no longer hold.

Simulations were also performed using the SRD method to test the effect, if any,
of hydrodynamics on the hopping process. We simulated a total of $1000$ freely
diffusing colloids in $2d$ pipes with widths characterized by $\delta_c \approx 0.035-0.093$
The longitudinal colloid density was $\eta=0.5$ in all cases.   A separate set of BD simulations was also performed so that the effect of HI could be clearly contrasted and compared.

The top panel of  Fig.~\ref{Fig:thopbdsrd2} shows that  the BD exhibits the same $\delta_c^{-2}$ scaling that was found in Fig.~\ref{Fig:thopdhop} for a higher packing fraction. For the larger values of $\delta_c$ shown, the BD and SRD simulations give almost the same value of $t_{hop}/t_D$.  However, for smaller $\delta_c$, the SRD simulations find a significantly higher value of $t_{hop}$, which suggests that the HI suppress hopping events.   We attribute this to the following processes.  When two particles come close together, or close to a wall, the solvent must be displaced for the particle to move past.  At very short distances, this gives rise to so-called lubrication forces~\cite{Happ73}.  These are repulsive for two particles approaching each other, and attractive when two particles are close together and then move away from each other (because now solvent has to flow into the space between them).    SRD simulations reproduce the lubrication forces, even on distances considerably less than $a_0$~\cite{Padd06}.   We thus attribute the increase of $t_{hop}$ with respect to the Brownian simulations to  lubrication forces, since they make it harder for two particles that diffuse towards each other to pass when $\delta_c$ is very small.

The bottom panel of Fig.~\ref{Fig:thopbdsrd2} shows the effective diffusion
coefficient plotted against the hopping time for the four largest separations.  Within simulation errors, we recover the  $D_{hop}\sim t_{hop}^{-1/2}$ scaling that was also observed for the BD simulations.  
Although the scaling law will break down for larger values of $\delta_c$, Figs.~\ref{Fig:thopdhop} and~\ref{Fig:thopbdsrd2} suggest that $D_{hop}$ will reach a value close to $D_0$ well before $\delta_c = 1$.  The exact value of $\delta_c$ where  $D_{hop}$ fully converges to that of a bulk system will of course depend on packing fractions and other details of the system.

\section{Conclusion}

We have carried out computer simulations to investigate single file diffusion and the crossover to Fickian diffusion when the particle passing constraint is lifted.   By comparing BD and SRD simulations, we can study the effect of HI on these processes.   Whereas HI have an important effect on the single particle diffusion coefficient under confinement, especially in 2d~\cite{Sane09a}, given $D_0$, we were unable to measure any further effects on the SFD mobility $F$ for particles that cannot pass each other.  The value we measure for $F$  at different packing fractions $\eta$ is consistent with the exactly solvable hard rod model~\cite{Harr65,Levi73}.   Experiments on colloidal suspensions~\cite{Wei00,Lin02,Lin05}, also find   $\lb x^2 \rb \approx 2 F t^\frac12$ scaling and in one case~\cite{Lin05} quantitative agreement with $F$ from the hard-rod model,  so simulations and experiment both suggest that HI have at best a very small effect on the SFD mobility.

When the no-passing constraint is lifted, we can measure the average hopping time $t_{hop}$ that it takes a particle to switch order with a neighbour.   While at shorter times $t_x \lesssim t \ll t_{hop}$ the particles still exhibit SFD, on time scales $t \gtrsim t_{hop}$, Fickian diffusion emerges with $\lb x^2 \rb \approx 2 D_{hop} t$.  The effective hopping diffusion coefficient scales as $D_{hop} \sim t_{hop}^{-\frac12}$ for small $\delta_c$. 

We find that for the Brownian simulations $t_{hop} \sim \delta_c^{-2}$ for small $\delta_c$,  as predicted by TST, but when HI are included, a stronger dependence becomes evident at small $\delta_c$.  We attribute this increase of $t_{hop}$ to the influence of hydrodynamic lubrication forces that make it harder for the particles to pass one another.

The hopping time $t_{hop}$ and diffusion coefficient $D_{hop}$ depend strongly on $\delta_c$.  This means that measuring how $t_{hop}$ or $D_{hop}$ depend on $\delta_c$ may be difficult for experiments, because very small relative errors in the channel width can lead to large effects.   But this sensitivity to changes in $\delta_c$ could also be used as an advantage, since it could be exploited, for example,  by microfluidic applications~\cite{Squi05}  as a  very sensitive way to  measure of the size of particles in artificial channels.

\acknowledgments
J.S. thanks Schlumberger Cambridge Research and IMPACT FARADAY for an
EPSRC CASE studentship which supported this work. A.A.L. thanks the Royal Society (London),
J.T.P. thanks the Netherlands Organisation for Scientific
Research (NWO) for financial support.
We thank L. Bocquet for helpful conversations.

\end{document}